\documentclass[superscriptaddress,twocolumn,showpacs,prb]{revtex4-1}
\usepackage[utf8]{inputenc} 
\usepackage{mathtools}
\usepackage{amsmath}
\usepackage{braket}
\usepackage{graphicx}
\usepackage{pgfplots}
\pgfplotsset{compat=1.15}
\usepackage{csquotes}
\usepackage{hhline}
\usepackage{tabularx}
\usepackage{subcaption}
\usepackage{amssymb}

\usepackage{dcolumn}
\usepackage{tabularx}
\usepackage[colorlinks=true,linkcolor=blue,citecolor=blue,urlcolor=blue]{hyperref}
\usepackage{braket}
\usepackage{float}
\usepackage{listings}
\usepackage{color}
\usepackage{subcaption}
\usepackage{placeins}
\definecolor{mygreen}{rgb}{0,0.6,0}
\definecolor{mygray}{rgb}{0.5,0.5,0.5}
\definecolor{mymauve}{rgb}{0.58,0,0.82}

\newcolumntype{C}{>{\centering\arraybackslash}X}

\begin{document}
\title{Simulating the Hamiltonian of Dimer Atomic Spin Model of One Dimensional Optical Lattice on Quantum Computers}

\author{Sudev Pradhan}
\email{sudev18@iiserbpr.ac.in}

\author{Amlandeep Nayak}
\email{amlandeep18@iiserbpr.ac.in}

\author{Sritam Kumar Satpathy}
\email{sritam19@iiserbpr.ac.in}

\author{Tanmaya Shree Behera}
\email{tanmaya18@iiserbpr.ac.in}

\affiliation{Department of Physical Science, \\Indian Institute of Science Education and Research, Berhampur, 760010 , Odisha, India}

\author{Ankita Misra}
\email{ankita18@iiserbpr.ac.in}

\author{Debashis Swain}
\email{debashis18@iiserbpr.ac.in}

\affiliation{Department of Chemical Science, \\Indian Institute of Science Education and Research, Berhampur, 760010 , Odisha, India}

\author{Bikash K. Behera}
\email{bkb18rs025@iiserkol.ac.in}
\affiliation{Department of Physical Sciences,\\ Indian Institute of Science Education and Research Kolkata, Mohanpur 741246, West Bengal, India}

\begin{abstract}
\begin{center}
Abstract
\end{center}
The one-dimensional Ising model with its connections to several physical concepts plays a vital role in comprehension of several principles, phenomena and numerical methods. The Hamiltonian of a coupled one-dimensional dissipative spin system in the presence of magnetic field can be obtained from the Ising model. We simulate the above Hamiltonian by designing a quantum circuit with precise gate measurement and execute with the IBMQ experience platform through different $N$ states with controlled energy separation where we can check quantum synchronization in a dissipative lattice system. Our result shows the relation between various entangled states, the relation between the different energy separation ($\omega$) with the spin-spin coupling ($\lambda$) in the lattice, along with fidelity calculations for several iterations of the model used. We also estimate the ground and first excited energy states of Ising-Hamiltonian using VQE algorithm and investigate the lowest energy values varying the number of layers of ansatz.
\end{abstract}


\maketitle
\section{Introduction}
Quantum advantage refers to the ability of quantum computers to solve problems at a faster rate than classical computers while utilizing reduced resources. In the 21st century there has been great progress towards establishment and progress of quantum advantage through demonstration of Shor's algorithm \cite{1,47} and implementation of Deutsch's algorithm \cite{2,48,49} through use of clustered quantum computers. However, the complexity required to build a quantum computer retaining all the advantages granted by quantum computation is impossible to attain with current technology. Hence, several sub universal models which retain parts of total advantage of quantum computation have been realized, use of which are widespread. These sub universal models are assigned to specific problems which are difficult to solve utilising classical machines, thereby demonstrating quantum advantages on several different aspects.

The quantum spin systems and the study of its lower energy model could be done by the ultra-cold atoms \cite{3,50} and this spin systems help us to explore more about the quantum phase transition, entanglement and other many body theoretical models \cite{4,51,52,53}. In order to get the desired effects or phenomenon in spin systems we need to minimise the degree of freedom of the coupling state, which would otherwise result collective dissipation and lead to destruction of quantum states. We have consider dimer atomic bosonic species with one dimensional optical lattice because this ultra-cold bosonic species serves best to simulate the Hamiltonian of the Ising spin model \cite{5,54,55,56,57,58}. Here we use Ising Model equivalent of a dimer atomic lattice  to learn of its Hamiltonian and its eigenvalue solutions\cite{60}. The Ising model of today is concerned with the physics of phase transitions, which occur when a small change in a parameter such as temperature or pressure causes a large-scale, qualitative change in the state of a system. Ising model starts with equally spaced points in a lattice of any dimension where each point has a spin of $\frac{1}{2}$ or $\frac{{-1}}{2}$. These points are independent and show coupling interaction due to mean/average fields of neighbouring spins. One purpose of the Ising model is to explain how short-range interactions\cite{59} between, say molecules in a crystal give rise to long-range, correlative behavior, and to predict in some sense the potential for a phase transition \cite{6}. The Ising model has also been applied to problems in chemistry \cite{7}, molecular biology \cite{8}, and other areas where ``cooperative" behavior of large systems is studied. Phase separation is one of the first facets where Ising model was used, it investigated spontaneous magnetization in ferromagnetic film (i.e. magnetization in the absence of external magnetic field) Transition temperature depends on the strength of the inter-spin exchange coupling, concepts of Ising model help in studying phase separation in binary alloys \cite{9} and liquid-gas phase transitions \cite{10}.

Solving the Hamiltonian in this 1D optical lattice would also help us to study quantum synchronisation. We understand Synchronization as ``adjustment of rhythms of the oscillating object due to their weak interaction" \cite{11} but in quantum state synchronizing a clock \cite{12} with quantum variables when the information is passed to two different person which might have different phase and state \cite{13} or the reduction of noise. This synchronisation is useful in steady super radiant emission \cite{14,15} in the harmonic oscillator. We can see this quantum synchronization in these atomic systems with the help of collective gratification coupling with the atoms and all because of super-radiance \cite{16} of incongruous effects. 

To solve this dissipative Ising-Hamiltonian and getting its eigenvalue along with its energy levels we use Variational quantum eigensolver (VQE). The VQE method was introduced to mitigate the significant hardware demands needed by the QPE (quantum phase estimation) approach. VQE is a hybrid quantum-classical algorithm, where the computational workload is shared between available classical and quantum resources \cite{17}. It starts with a reasonable assumption about the form of the target wave function with the ground state trial wave function being generated from operators that result in single and double-excitation configurations from Hartree-Fock wave function \cite{18} which is precomputed on a classical computer. Next, on a quantum computer, the trial wave function is prepared and the expectation value of the Hamiltonian is measured. Then, the parameters of the trial wave function are optimized iteratively on a classical computer using the variational principle \cite{19}.

VQE \cite{32} provides eigen-solutions for trail ground state and first excited state wavefunctions.The ground and excited state energy are then plotted through several graphs to check their general trend, their dependency on several individual operators used as gates and to check a general maxima or minima of the energy eigenvalues.

In the following sections we show how to apply the prior mentioned portions. Specifically, we show how to implement Ising-Hamiltonian on IBM Quantum computer by setting up intermediate circuit and performing several observations on different relations such as that of:
\begin{itemize}
 \item Time correlation and average value using different values for spin spin interaction term \cite{33}/coupling constant.
 \item Different energy separations and coupling constant.
\end{itemize}

Fidelity calculation \cite{34} for the circuit is also done by using the data obtained by computing the circuit both classically and with quantum computers.
Eigen-solutions for the Ising-Hamiltonian is also obtained using VQE, we have also included several plots displaying the relation (i) of energy with number of layers, and (ii) of energy with changing external field intensity, in order to better explain various trends of the obtained energy. 

\section{Theory\label{qnm_Sec2}}

Super cold atom in atomic lattice $\cite{20}$ is perfect to simulate the Hamiltonian spin $\cite{21}$ complex where we can implement by two different techniques i.e. optical driving of two hyperfine levels $\cite{22}$ and interacting with the Ising spin $\cite{23}$ of mott-insulator \cite{24}. Using the later technique we simulate the Hamiltonian Operator with the effective spin of a single atom where they are restricted to each lattice potential and this lattice potential is localised and basic ground states and excited states are marked as $\Ket{0}$ and $\Ket{1}$ and a effective spin of $\frac{1}{2}$ in each lattice where these collective atoms work in a restricted region. With the use of lattice potential \cite{25} we can state the disintegration of atoms which ultimately end up, cooling of the system. When we simulate this spin system the ground state is achieved faster than the excited states and thus it will eliminate the excited states which will reduce to desired 2 level system. The simple Hamiltonian by the optical coupling is mentioned in the qualitative description in the paper \cite{26}. This sets up the equation for a two level system as:

\begin{eqnarray}\label{eqndissmain}
\dot{\rho}_t=&\sum_k A^-(2\sigma_k^-\rho_t
\sigma_k^+-\{\sigma_k^+\sigma_k^-,\rho_t\}_+)-i\left[H
,\rho_t\right].
\label{qs_Eq1}
\end{eqnarray}

where, $\sigma_k^- = ({\sigma_k^+})^{\dagger}$ , $\sigma_k^+ = \Ket{1}{\bra{0}_k}$ and $A^-$ indicates the terms of decay as mentioned in $\cite{26}$.

And using this main equation from this $\cite{27}$ paper and considering all level of the atomic lattice and coupled through a detunned Raman transition \cite{28} by passing through the excited state and we obtain the Fourier integral of the Hamiltonian states $\cite{29}$ containing discrete staggered energy states and the spin spin coupling ($\lambda$).

Assuming the on-site interaction, a single potential well consist of 2 atoms, where we can use the perturbation theory. Implementing this dissipative model technique and double well we can reduce the Hamiltonian Operator to a summation of staggered energy of the lattice dimer with  the coupling complex and magnetic field in z-direction and tunable transverse field. In this XXZ spin chain model, if we don't consider the z-direction of magnetic field, it would effectively calculate the steady state dynamics. And we consider the time dependent equation for the x direction of magnetic field for the equation $\eqref{qs_Eq1}$ as mentioned in the dissipative spin chain section of \cite{26} so using the perturbation theory and the given manipulation the optical well and we can sum up the Hamiltonian as follows:- \cite{27}
\begin{equation}\label{Ham}
\hat{H}=\sum_{j=1}^{N}\frac{\omega_j}{2}\hat{\sigma}^z_j+ 
\sum_{j=1}^{N-1} \lambda(\hat{\sigma}^+_j\hat{\sigma}^-_{j+1}+h.c.), 
\end{equation}\\

where $\omega_0\gg\omega_{1,2}, \delta, \lambda$, where $\delta=\omega_1-\omega_2$ \cite{1} is the detunned coefficient of the lattice potential and ($\hbar=1$).

In order to reduce the Hamiltonian into a quantum circuit, we take the simplest state for $N=2$, (2 qubit system), and construct the circuit. Similarly for the next iteration of N, it follows the same trend.
We apply the time evolution unitary operator \cite{30} on our Hamiltonian to get,
\begin{eqnarray}
    U(t)=e^{-\frac{i\hat{H}t(2\pi)}{h}}
\label{Ham_3}
\end{eqnarray}
where $t$ refers to time and $\hat{H}$ is the combination of $H_1$ and $H_2$, where $H_1 = \sum_{j=1}^{N}\frac{\omega_j}{2}\hat{\sigma}^z_j$ and $H_2 = \sum_{j=1}^{N-1} \lambda(\hat{\sigma}^+_j\hat{\sigma}^-_{j+1}+h.c.)$ but due to the presence of $\lambda(\sigma_x\sigma_x + \sigma_y\sigma_y)$, it does not allow these two Hamiltonian to commute, $[H_1, H_2] \neq 0$    $\cite{31}$, hence we use totem decomposition  
and by taking $\frac{h}{2\pi}=1$ we could write the equation as 

\begin{eqnarray}
    U(t)=[(e^{-\frac{i{H_1}t}{n}}).(e^{-\frac{i{H_2}t}{n}})]^n\nonumber\\
\label{Ham_111}
\end{eqnarray}
where $\frac{t}{n}$ is the small interval which we can write as $\Delta t$ and our above equation will be modified to 
\begin{eqnarray}
    U(t)=[(e^{i{H_1}\Delta t}).(e^{i{H_2}\Delta t})]^n\nonumber\\
\label{Ham_121}
\end{eqnarray}
where we will repeat the process n times or theoretically we will iterate our quantum circuits n number of times. As the value of n increases we will get more accurate results. Now substituting the value of $H_1$ and $H_2$ in the above equation $\eqref{Ham_121}$, we get 
\begin{eqnarray}
    U(t)=[(e^{i{(\frac{\omega_j}{2}\hat{\sigma}^z_j)}\Delta t}).(e^{i{\lambda(\hat{\sigma}^+_j\hat{\sigma}^-_{j+1}))}\Delta t})]^n\nonumber\\
\label{Ham_4}
\end{eqnarray}

where, $\sigma^x$= $\begin{bmatrix}
  0 & 1\\ 
  1 & 0
\end{bmatrix}$, $\sigma^y$= $\begin{bmatrix}
  0 & -i\\ 
  i & 0
\end{bmatrix}$ and $\sigma^z$= $\begin{bmatrix}
  1 & 0\\ 
  0 & -i
\end{bmatrix}$\\ \cite{35}
Dividing the equation \eqref{Ham_4} into 2 parts as -

\begin{eqnarray}
    \label{Ham_131}  U_a(t)&=&[e^{i{(\frac{\omega_j}{2}\hat{\sigma}^z_j)}\Delta t}]^n\\\label{Ham_141} \nonumber\\ 
    U_b(t)&=&[e^{i{\lambda(\hat{\sigma}^+_j\hat{\sigma}^-_{j+1})}\Delta t}]^n\\ 
\nonumber
\end{eqnarray}

\begin{figure}[H]
    \centering
    \includegraphics[width=0.5\textwidth]{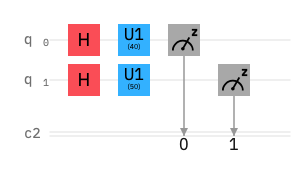}
    \caption{Circuit describing $H$ gate with $U1$ gate on q[0] qubit where $U1$ gate has $\omega_1 t$ as $\frac{\pi}{2}$ where we get a entangled state.}
    \label{fig1}
\end{figure}

On solving the equation \eqref{Ham_131} (the first part of the equation) and assuming for a single iteration (n=1) and this follows the condition of \cite{36}

\begin{eqnarray}
    e^{-i\theta A}=cos(\theta) I -isin(\theta)A\nonumber\\
\label{Ham_151}    
\end{eqnarray}

where $I$ is the identity matrix and $A$ is of same order of the identity matrix where $A^2 = I$. Here, $A=\sigma_z $, which follow the identity rule($\sigma_z^2 =I $) which has a order 2, and $\theta = \frac{\omega_j \Delta t}{2}$, substituting this, we get
\begin{eqnarray}
    U_a(t)&=&cos(\frac{\omega_j \Delta t}{2})I-isin(\frac{\omega_j \Delta t}{2})\sigma_z\nonumber\\
\label{Ham_161}
\end{eqnarray}
under matrix transformation,

\begin{equation}
    U_a(t)=\begin{bmatrix} e^\frac{-i\omega_j \Delta t}{2} & 0\\ 0 & e^\frac{i\omega_j \Delta t}{2} \end{bmatrix} 
    = \exp^\frac{-i\omega_j \Delta t}{2}\begin{bmatrix}1 & 0\\0 & e^{i\omega_j \Delta t}\end{bmatrix}
\label{Ham_171}    
\end{equation}
this $U_a(t)$ matrix represent a similar matrix of $U1$ gate on IBMQ where $U1$ ($\theta t$) is replace by $U1$ ($\omega_j \Delta t$) and according to our Hamiltonian $\omega$ is flexible. So applying $U1$ gate on first qubit satisfy j=1 term where $U1$ ($\omega_1 \Delta t$) and  same goes with the another qubit for j=2, we get another $U1$ ($\omega_2 \Delta t$) gate and goes on till N. But in order for the $U1$ gate to work we need a Hadamard gate in order to produce a superposition state which will activate the $U1$ gate as in Fig: \ref{fig1}

Similarly, working for the equation \eqref{Ham_141} we get a reduced matrix format which on solving the later part of the equation, $e^{i\lambda(\hat{\sigma}^+_j\hat{\sigma}^-_{j+1} \Delta t)}$, with considering as $\sigma^+=\sigma_x+i\sigma_y$ and $\sigma^-=\sigma_x-i\sigma_y$ which reduce down the equation to $e^{\lambda(\sigma_x\sigma_x + \sigma_y\sigma_y)\Delta t}$ or can be written as $e^{\lambda(\sigma_x\sigma_x)\Delta t} e^{(\sigma_y\sigma_y)\Delta t}$and solving the $e^{\lambda\Delta t(\sigma_x\sigma_x)}$ in a Euler formula $\cite{36}$, we get

\begin{eqnarray}
    e^{-i\theta B}=cos(\theta) I -isin(\theta)B\nonumber\\
\label{Ham_181}    
\end{eqnarray}
where $I$ is the identity matrix and $B$ is of same order of the identity matrix.\\
Here, B=$\sigma_x \otimes \sigma_x$, which follow the identity rule and has a order 4 and $\theta = \lambda\Delta t $, substituting this, we get
\begin{eqnarray}
    U_1(t)&=&cos(\lambda\Delta t)I-isin(\lambda\Delta t)\sigma_x \otimes \sigma_x\\\nonumber
\label{Ham_191} 
\end{eqnarray}

\begin{figure}[H]
\centering
\includegraphics[width=0.5\textwidth]{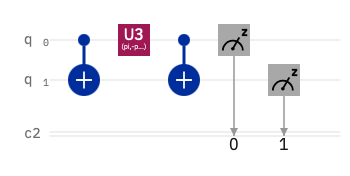}
\caption{\textbf{The derived circuit has a combination of a CNOT gate, $U3$ gate and a CNOT gate, where $\theta=\lambda \Delta t$, $\phi=-\pi/2$ and $\lambda=\pi/2$}.}
\label{fig2}
\end{figure}

And converting to the matrix format, 

\begin{equation}
\begin{bmatrix}
cos(\lambda \Delta t) & 0 & 0 & -isin(\lambda \Delta t)\\
0 & cos(\lambda \Delta t) & -isin(\lambda \Delta t) & 0\\
0 &  -isin(\lambda \Delta t) & cos(\lambda \Delta t) & 0\\
-isin(\lambda \Delta t) & 0 & 0 & cos(\lambda \Delta t)
\end{bmatrix}
\end{equation}

which shows a normal $U3$ ($\theta,-\frac{\pi}{2},\frac{\pi}{2}$) on IBMQ
where $\theta$ varies according to $\lambda \Delta t$ vary.
Where $U3$ gate controls The three parameters allowing the construction of any single-qubit gate, has a duration of one unit of gate time. In the Bloch sphere rotation $\cite{37}$, it can move through any plane controlled by ($\theta$, $\gamma$ and $\phi$) only once and its matrix form is represented by \\
\begin{equation}
\begin{bmatrix}
   cos(\frac{\theta}{2}) & -e^{i\lambda}sin(\frac{\theta}{2})\\ 
   e^{i\phi}sin(\frac{\theta}{2}) & e^{i(\phi+\lambda)}cos(\frac{\theta}{2})
\end{bmatrix}\nonumber\\
\end{equation}
hence putting two CNOT gates i.e.\\
\begin{equation}
\begin{bmatrix}
   1 & 0 & 0 & 0\\ 
   0 & 1 & 0 & 0\\
   0 & 0 & 0 & 1\\
   0 & 0 & 1 & 0
\end{bmatrix}\nonumber\\
\end{equation}
and $U3$ with defined parameter in between, we can obtain the result of the above matrix of $U1$.
Hence, solving the matrix and comparing it with the $U3$ matrix we derive the following circuit as in Fig: \ref{fig2}:\\
Similar the second part of \eqref{Ham_141}, $\lambda\Delta t(\sigma_y\sigma_y)$ we could calculated as,\\

\begin{eqnarray}
    U_b(t)&=&cos(\lambda\Delta t)I-isin(\lambda\Delta t)\sigma_y \otimes \sigma_y\nonumber\\
\nonumber
\end{eqnarray}
\begin{equation}
\begin{bmatrix}
cos(\lambda \Delta t) & 0 & 0 & isin(\lambda \Delta t)\\
0 & cos(\lambda \Delta t) & -isin(\lambda \Delta t) & 0\\
0 &  -isin(\lambda \Delta t) & cos(\lambda \Delta t) & 0\\
isin(\lambda \Delta t) & 0 & 0 & cos(\lambda \Delta t)
\end{bmatrix}
\end{equation}
similarly on solving this matrix manually, we get a combination of control $U3$ and anti control $U3$ matrix, where control $U3$ is a 4 $\times$ 4 matrix ($I \otimes$ $\ket{0}\bra{0}$ + $U3 \otimes$  $\ket{1}\bra{1}$) i.e.\\
\begin{equation}
\begin{bmatrix}
   1 & 0 & 0 & 0 \\ 
   0 & cos(\frac{\theta}{2}) & 0 & -e^{i\lambda}sin(\frac{\theta}{2}) \\
   0 & 0 & 1 & 0\\
   0 & e^{i\phi}sin(\frac{\theta}{2}) &  0 & e^{i(\phi+\lambda)}cos(\frac{\theta}{2})
\end{bmatrix}\nonumber\\
\end{equation}
and anti control $U3$ is also a 4$\times$4 matrix ($I \otimes$ $\ket{1}\bra{1}$ + $U3 \otimes$  $\ket{0}\bra{0}$) i.e.\\
\begin{equation}
\begin{bmatrix}
   cos(\frac{\theta}{2}) & 0 & -e^{i\lambda}sin(\frac{\theta}{2}) & 0 \\ 
   0 & 1 & 0 & 0 \\
    e^{i\phi}sin(\frac{\theta}{2}) & 0 & e^{i(\phi+\lambda)}cos(\frac{\theta}{2}) & 0\\
   0 & 0 & 0 & 1
\end{bmatrix}\nonumber\\
\end{equation}
and placing this combination of matrix within 2 CNOT matrix with specified value of ($\theta$, $\gamma$, $\phi$) we get the above matrix.\\

\begin{figure}[H]
\centering
\includegraphics[width=0.5\textwidth]{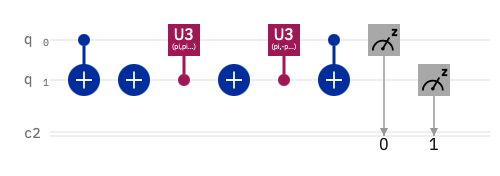}
\caption{\textbf{The derived circuit has a combination of a CNOT gate, control $U3$ gate, Anti-control $U3$ gate and a CNOT gate, where $\theta_1=\lambda \Delta t$, $\phi_1=-\pi/2$ and $\lambda_1=\pi/2$ and $\theta_2=\lambda \Delta t$, $\phi_2=\pi/2$ and $\lambda_2=-\pi/2$}.}
\label{fig3}
\end{figure}

\begin{figure}[H]
    \centering
    \includegraphics[width=0.5\textwidth]{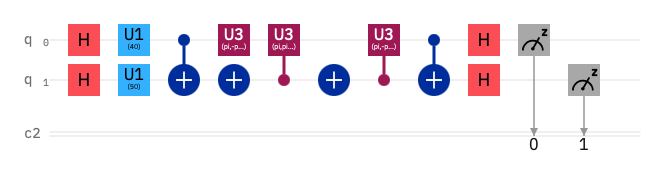}
    \caption{\textbf{The equivalent circuit describes the Hamiltonian operator \ref{Ham} in a N=2 state.}}
    \label{fig4}
\end{figure}

Where the equivalent matrix is reduced in the form of circuit, as per Fig: \ref{fig3}.\\

and summing up all the circuits for N=2, we can add all the circuits together and as we know that two CNOT gates are equal to identity matrix so we can omit that and the resulting circuit somewhat looks like this as in Fig: \ref{fig4}.\\

We can say the following matrices are the Ising matrices \cite{38} which confirm the Ising spin complex of the atom in the lattice potential.\\
$\vspace{15mm}$

\section{Implementation on IBM Quantum Experience \label{qnm_3}}
\subsection{Setting up the intermediate circuit}
$\vspace{3mm}$

The primary task is to implement the above constraint as the unitary time evolution operator \cite{30}, which we have done through implementing through $U1$, $U3$ and $U3^\dag$ 
\begin{figure}[H]
    \centering
    \includegraphics[scale=0.18]{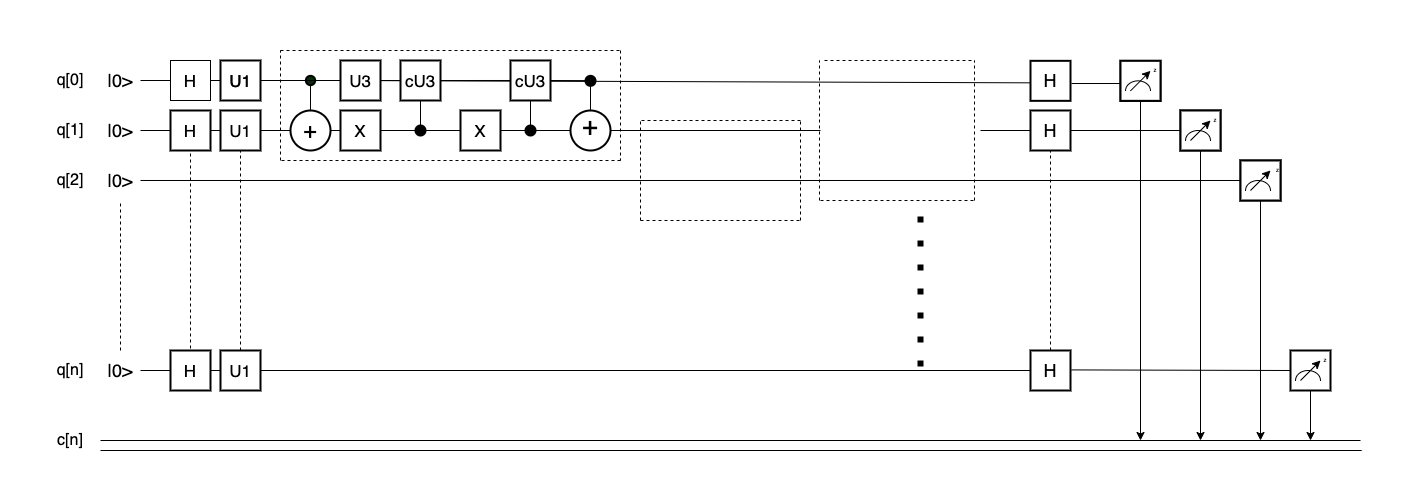}
    \caption{The skeletal body of the general circuit for N qubits.}
    \label{fig5}
\end{figure}

gates by controlling their measures and converting them to the basic quantum circuit so that the phase is not disturbed. We set the parameters $\theta$, $\phi$ and $\lambda$ as per our requirement to simulate $\hat{H}$. \\

Setting the \eqref{qnm_3} as a basic circuit on IBM Q as implementing the simple $U1$ gate along with the measurement gate, If we choose $\theta$ as $\frac{\pi}{2}$, q[0] gives $\Ket{0}$ and q[1] gives $\Ket{1}$ and we use the Hadamard gate along with $U1$ gate we get a superposition of the two states $\Ket{0}$ $\Ket{1}$ as in fig:$\ref{fig1}$. \\
In the above equation $\eqref{Ham}$ with the precise measurement $\omega_1 t$ we can get the synchronized state.  \\

\begin{figure}[H]
\begin{subfigure}{0.465\textwidth}
\includegraphics[width=0.96\linewidth]{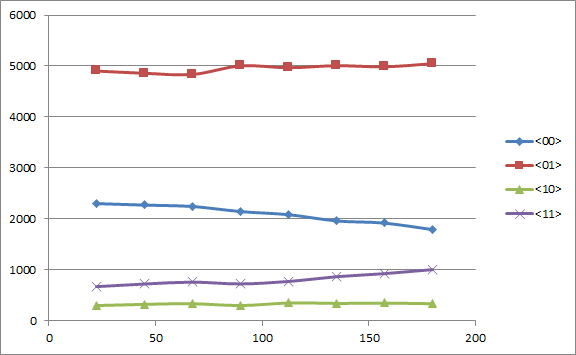} 
\caption{}
\label{fig:subim2}
\end{subfigure}\hfill
\begin{subfigure}{0.45\textwidth}
\includegraphics[width=\linewidth]{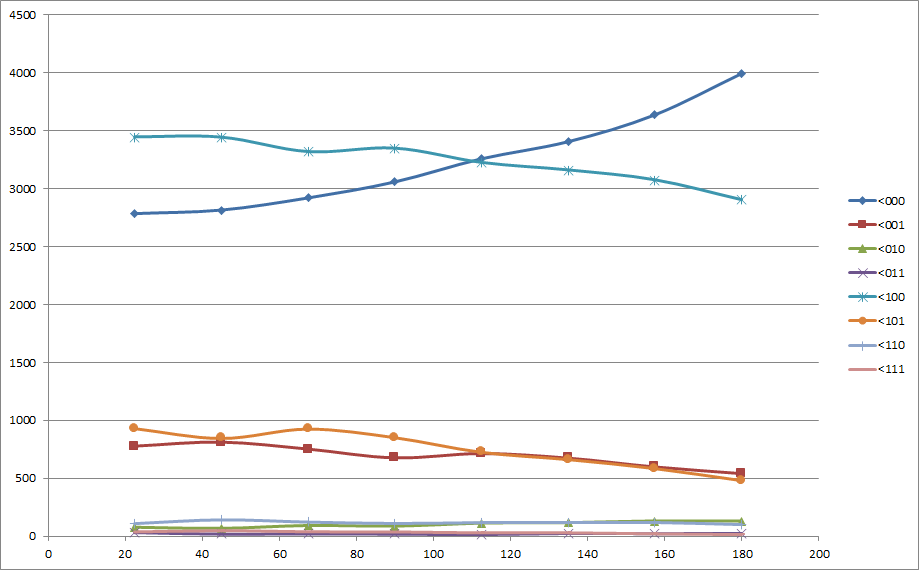} 
\label{fig:subim1}
\caption{}
\end{subfigure}\hfill
\begin{subfigure}{0.45\textwidth}
\includegraphics[width=\linewidth]{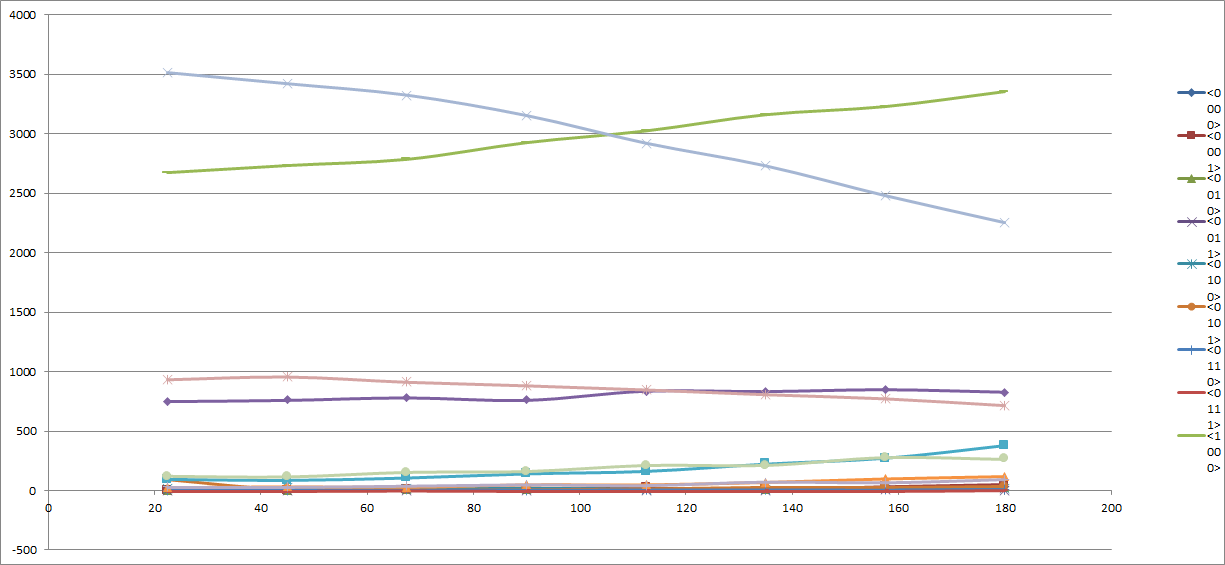} 
\caption{}
\label{fig:subim2}
\end{subfigure}\hfill
\caption{\textbf{Simulation results for probabilities of each state.}(a) When N=2 (b) N=3 (c) N=4 and (d) N=5; it shows the variation in probability of each state with respect to $\lambda t$, where $\lambda t$ is in term of degree radian and the y axis shows the percentage probability and each line indicate each entangled state with N qubits.}
\label{fig6}
\end{figure}

\begin{figure}[H]
\begin{subfigure}{0.48\textwidth}
\includegraphics[width=0.96\linewidth]{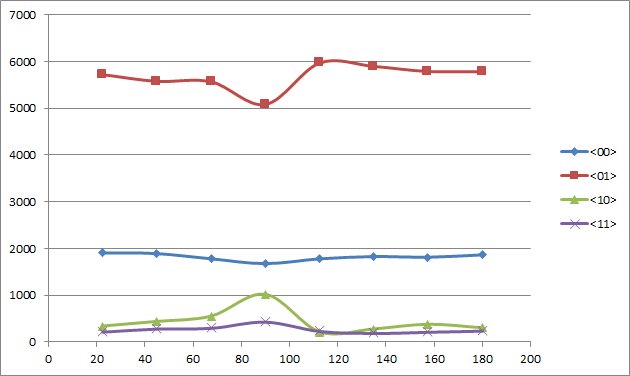} 
\caption{}
\label{fig:subim2}
\end{subfigure}\hfill
\begin{subfigure}{0.455\textwidth}
\includegraphics[width=\linewidth]{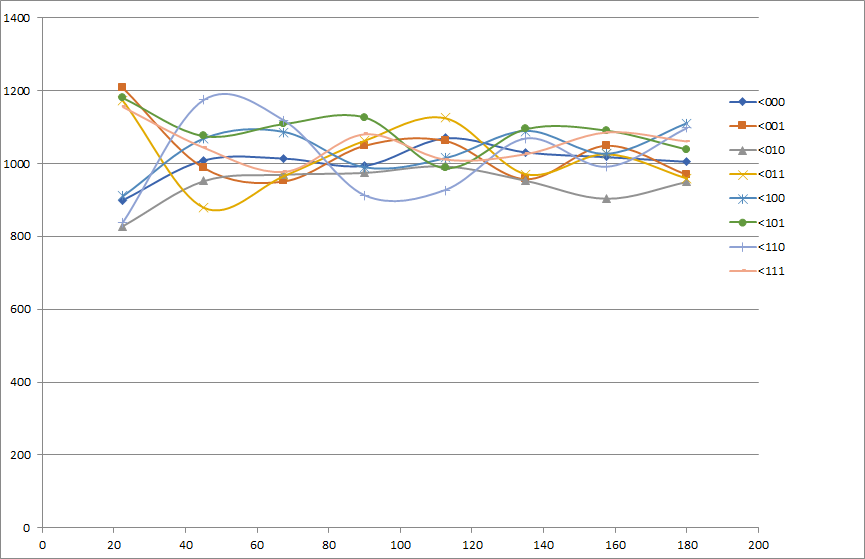} 
\label{fig:subim1}
\caption{}
\end{subfigure}\hfill
\begin{subfigure}{0.455\textwidth}
\includegraphics[width=\linewidth]{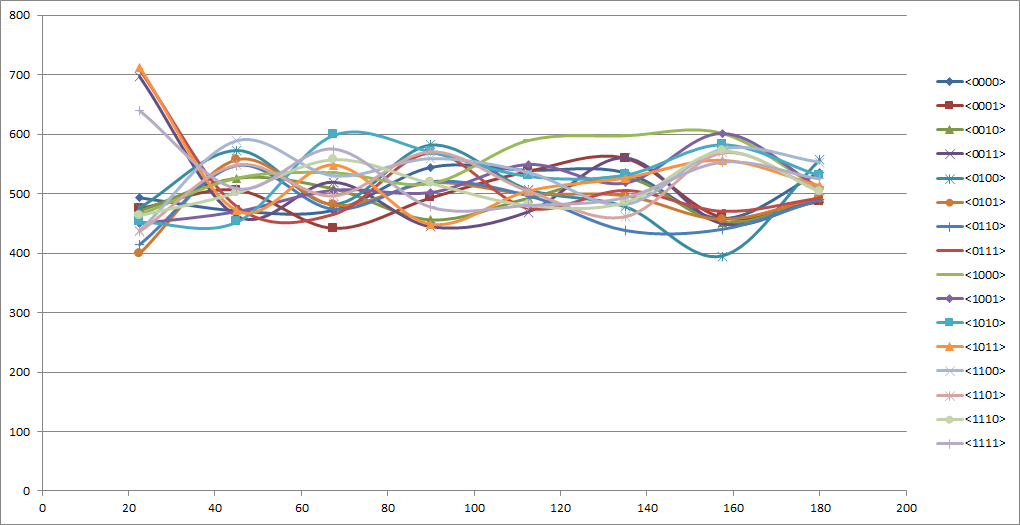} 
\caption{}
\label{fig7}
\end{subfigure}\hfill
\caption{\textbf{Quantum computer results for probabilities of each state.}(a) When N=2 (b) N=3 (c) N=4 and (d) N=5; it shows the variation in probability of each state with respect to $\lambda t$, where $\lambda t$ is in term of degree radian and the y axis shows the percentage probability and each line indicate each entangled state with N qubits.}
\label{qdctc_Fig3}
\end{figure}

In the Fig: $\ref{fig6}$ and $\ref{fig7}$ is for N=2 showing the percentage of each state in both classical simulators and real quantum machine respectively with the desired result which shows the entanglement.\\

Similarly, we set the second part of the equation $\eqref{Ham}$, by applying the unitary time evolution operator, we get $U3$ and $U3^\dag$ gates with precise measurement of $\theta$ ,$\phi$ and $\lambda$; where $\phi$ and $\lambda$ are $-\frac{\pi}{2}$ and $\frac{\pi}{2}$ respectively for $U3$ gate where as $\phi$ and $\lambda$ are $\frac{\pi}{2}$ and $-\frac{\pi}{2}$ respectively for $U3^\dag$ gate. Adding the following information we can reduce the quantum circuits as Fig: \ref{fig4} with the desired entangled state probability as in Fig: \ref{fig6} and \ref{fig7}. The general skeletal body of the Hamiltonian Operator with defined value of $\theta$, $\phi$ and $\lambda$ as in Fig: \ref{fig5}.\\

When simulating the Hamiltonian Operator for N states we get a result as in Fig: \ref{qdctc_Fig3} with each N state with varying $\lambda t$, which shows different entangled states for particular degree radian.\\

\subsection{Time correlation and the Average value}
When we take in option with two or more than two body system we are introduced by a factor called phase locking \cite{39} but this one-time correlation indeed help us to see the relationship between the average value and its weak dependence on $\lambda$ which is a condition for the synchronizing states. This average value somewhere  depends on particular eigenvalue or particular spin 

quantum number \cite{40}. While calculating the average value we can see the overlapping of the states which could be concluded as the opposite spin states. We observe the resonance peaks of the synchronized \cite{41} and the rest other eigenstates, where we can also observe the relation between $\lambda$ and $\omega$. 

To show this variation of weak dependence of spin quantum number on the average value and its dependence, we make a circuit that depicts the average value of each Pauli's matrices. These circuits also depict the relation between the $\lambda$ and $\omega$ for each excited state in the synchronization system. These synchronized states depend on many body or many-time correlation function, dependency on time is very low.
\begin{figure}[H]
    \centering
    \includegraphics[scale=0.3]{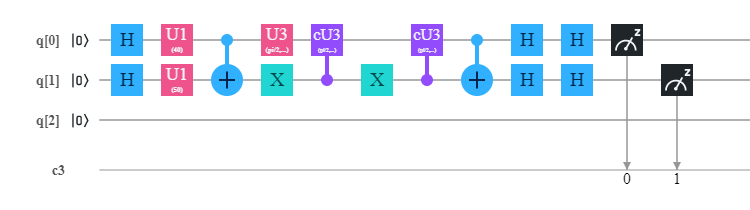}
    \caption{Circuit representation of the average value of $\braket{\sigma_x|\sigma_x}$ when N=2 state adding a $H$ gate at the end of the main circuit representing the operator.}
    \label{fig8}
\end{figure}
\begin{figure}[H]
    \centering
    \includegraphics[scale=0.3]{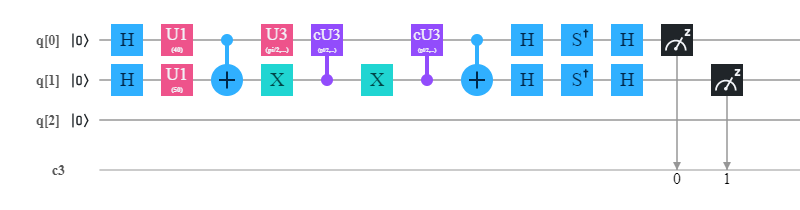})
    \caption{Circuit representation of the average value $\braket{\sigma_y|\sigma_y}$ when N=2 state keeping $S^\dag$ and $H$ gate at end.}
    \label{fig9}
\end{figure}
Here, we would show the circuit of the average value of $\braket{\sigma_x|\sigma_x}$ in Fig: \ref{fig8} by connecting another Hadamard gate.

at the end of the circuit with a particular $\theta$ as $\frac{\pi}{2}$ and rest other coefficient for the $U3$, $\phi$ and $\lambda$ as $-\frac{\pi}{2}$ and $\frac{\pi}{2}$ respectively and $U3^\dag$ ,$\phi$ and $\lambda$ as $\frac{\pi}{2}$ and $-\frac{\pi}{2}$ respectively.

Similarly, $\braket{\sigma_y|\sigma_y}$ are the average value which depicts the time correlation function. We can theoretically obtain this by showing the excitation state overlapping with the other eigen value, but in quantum circuit wise, we can show them by using a $S^\dag$ followed by a Hadamard gate at the end of the circuit Fig: $\ref{fig9}$ where $\theta$ as $\frac{\pi}{2}$ and rest other coefficient for the U3 and $U3^\dag$ gate remain the same.

\begin{figure}[H]
\begin{subfigure}{0.5\textwidth}
\includegraphics[width=\linewidth]{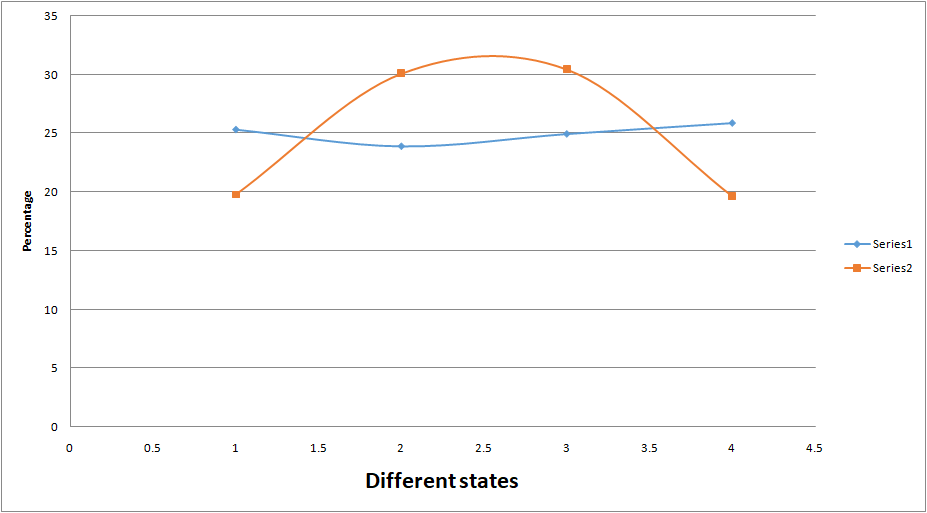} 
\label{fig:subim1}
\caption{}
\end{subfigure}\hfill
\begin{subfigure}{0.5\textwidth}
\includegraphics[width=\linewidth]{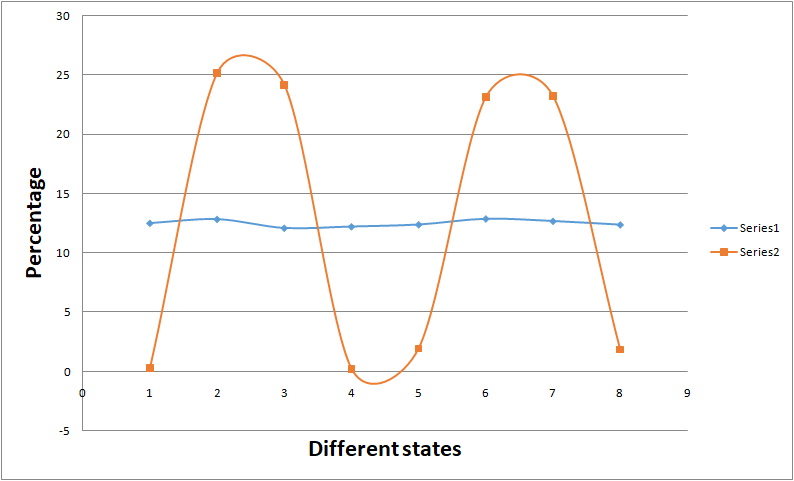} 
\caption{}
\label{fig:subim2}
\end{subfigure}\hfill
\begin{subfigure}{0.5\textwidth}
\includegraphics[width=\linewidth]{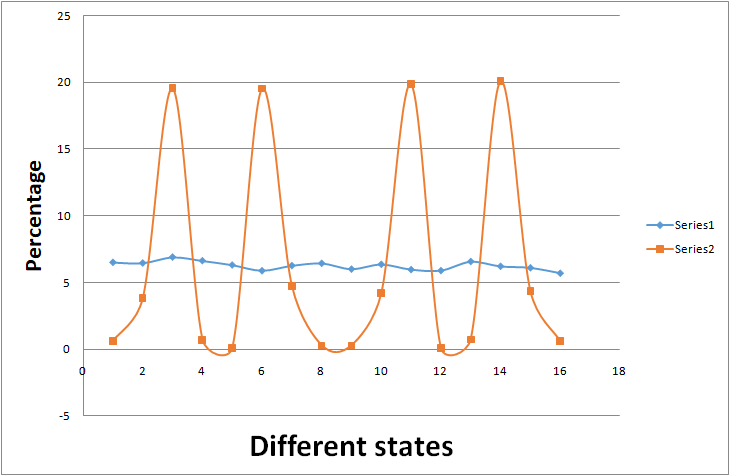} 
\label{fig:subim1}
\caption{}
\end{subfigure}\hfill
\caption{\textbf{Variation of probabilities of different states in x and y basis.}
(a) When N=2 (b) N=3 and (c) N=4; it shows the average value of each state (blue line representing $\braket{\sigma_x|\sigma_x}$ and red line representing $\braket{\sigma_y|\sigma_y}$) for particular $\lambda t$ = $\frac{\pi}{2}$ where we can check the overlapping value of the particular eigenstate and we could also check the resonance peak.}
\label{fig10}
\end{figure}

Merging all this information we could relate the data from Fig: $\ref{fig10}$, where we have simulated N=2 to N=4.

\subsection{Relation between the different energy separation ($\omega$) with the spin-spin coupling ($\lambda$)}

As per the dissipative state \cite{26}, we can have different $\omega$ (transverse field term) in Ising-Hamiltonian for different lattice points. This along with the spin spin coupling \cite{42} constant are necessary in order to complete the Hamiltonian. In order to check any relation between these terms we have varied $\theta$ values of either $U$ gate while keeping the other constant, the $\lambda$t thereby the theta for U3 gate is changed regularly irrespective of the changes in $\omega$. The results obtained is plotted for various  constant values of different $\omega$ in Fig: \ref{fig14} and \ref{fig15}.
Here, we find the synchronised state of the system for a particular $\omega$ and $\lambda$. From the provided plots, it can be easily observed that results due to $\omega_2$ remaining as constant is not much affected by varying $\lambda$t whereas results due to $\omega_1$ remaining as constant shows drastic change with varying $\lambda$t with the probabilities showing different change pattern for different states.

\section{Variational Quantum Eigensolver}
VQE \cite{32} is a hybrid module using both quantum and classical resources to find expectation values of $H$, where $H$ is the Hamiltonian of the entire system when run in quantum simulations such as in this case. For VQE a quantum subroutine is run inside of a classical optimization loop, in two fundamental steps:

\begin{itemize}
    \item Prepare the quantum state $\ket{\psi(\Vec{\theta})}$ often called then ansatz.
    \item Measure the expectation value $\bra{\psi(\Vec{\theta})}H\ket{\psi(\Vec{\theta})}$   
\end{itemize}

The trial quantum state/parametrized state (or ansatz) preparation can be tricky, the algorithms need to be considered carefully since it can dramatically affect performance. By arbitrarily selecting a wave function $\ket{\psi}$ (called an ansatz) as an initial guess, approximating it as $\ket{\psi_{min}}$, calculating its expectation value, $\bra{\psi} H \Ket{{\psi}}$ and iteratively updating the wave function, arbitrarily tight bounds on the ground state energy of a Hamiltonian may be obtained. The variational principle \cite{43} ensures that this expectation value is always greater than the smallest eigenvalue of $H$.\\

For this particular case we have used a simplified ansatz of $U3$ gates. A simplified form of it can be shown:
\begin{figure}[H]
\centering
\includegraphics[width=0.5\textwidth]{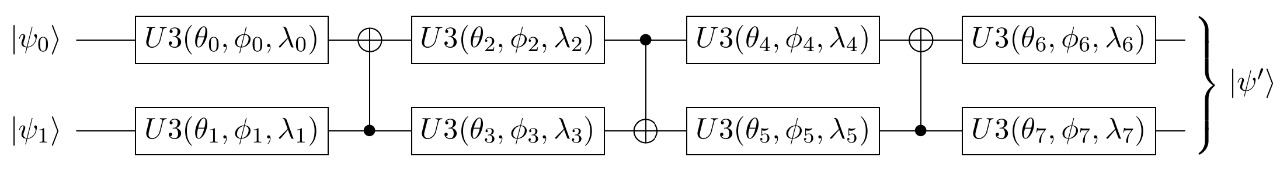}
\label{bhak}
\end{figure}

This ansatz is just limited to 3 parameters and hence can be efficiently optimized. It is to be understood that the capacity to create an arbitrary state guarantees that during the optimization process, the variational form

\begin{figure}[H]
\begin{subfigure}{0.38\textwidth}
\includegraphics[width=\linewidth]{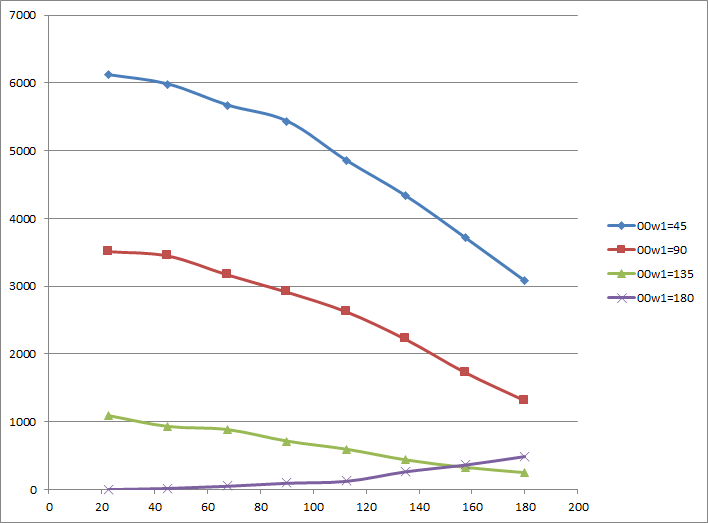} 
\label{fig:subim1}
\caption{\textbf{ `00' for constant values of $\omega_1$ }}
\end{subfigure}\hfill
\begin{subfigure}{0.38\textwidth}
\includegraphics[width=\linewidth]{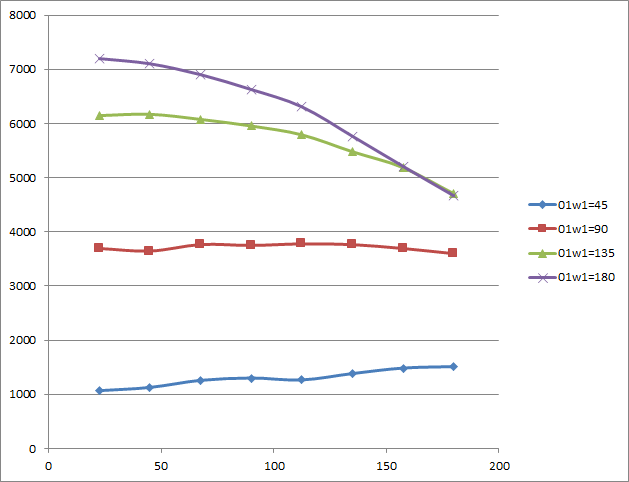} 
\caption{\textbf{`01' for constant values of $\omega_1$ }}
\label{fig:subim2}
\end{subfigure}\hfill
\begin{subfigure}{0.38\textwidth}
\includegraphics[width=\linewidth]{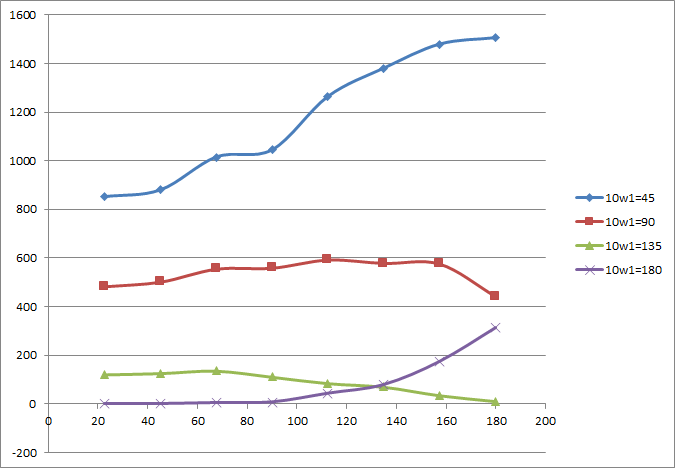} 
\caption{\textbf{`10' for constant values of $\omega_1$ }}
\label{fig:subim2}
\end{subfigure}\hfill
\begin{subfigure}{0.38\textwidth}
\includegraphics[width=\linewidth]{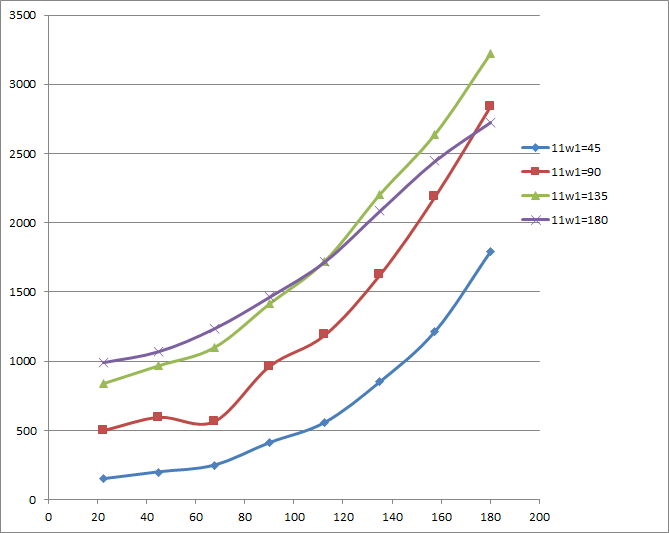} 
\caption{\textbf{`11' for constant values of $\omega_1$ }}
\label{fig:subim2}
\end{subfigure}\hfill
\caption{\textbf{Variation of probabilities for different states with $\lambda$ for constant values of $\omega_1$ while $\omega_2$ varies for N=2. }(a) `00' (b) `01' (c) `10' (d) `11'} 
\label{fig14}
\end{figure}

\begin{figure}[H]
\begin{subfigure}{0.37\textwidth}
\includegraphics[width=\linewidth]{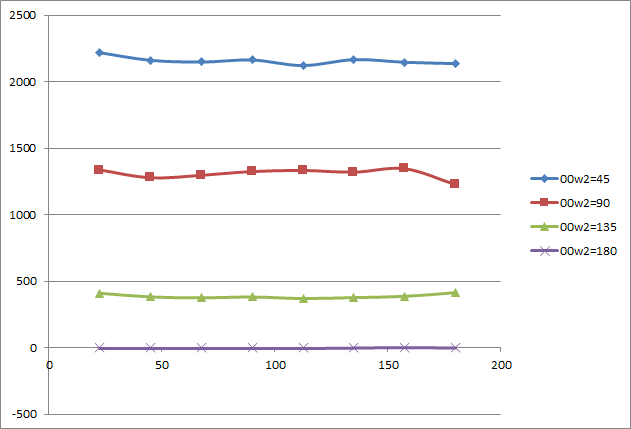} 
\label{fig:subim1}
\caption{`00' for constant values of $\omega_2$}
\end{subfigure}\hfill
\begin{subfigure}{0.37\textwidth}
\includegraphics[width=\linewidth]{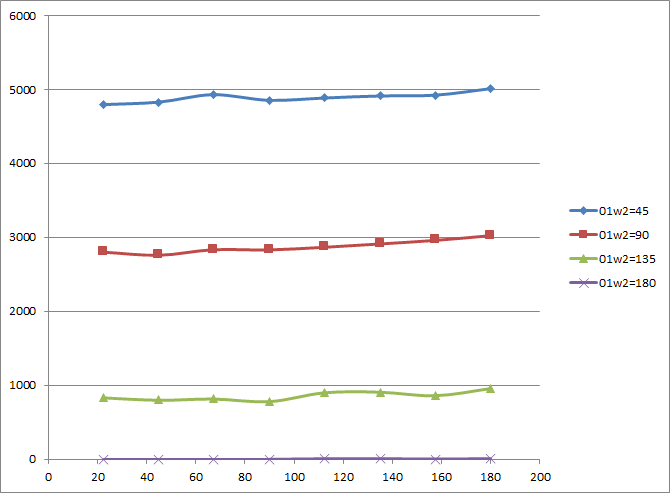} 
\caption{'01' for constant values of $\omega_2$}
\label{fig:subim2}
\end{subfigure}\hfill
\begin{subfigure}{0.37\textwidth}
\includegraphics[width=\linewidth]{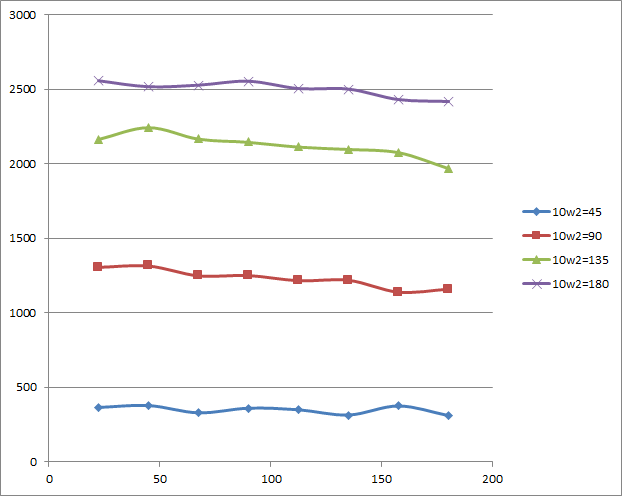} 
\caption{`10' for constant values of $\omega_2$}
\label{fig:subim2}
\end{subfigure}\hfill
\begin{subfigure}{0.37\textwidth}
\includegraphics[width=\linewidth]{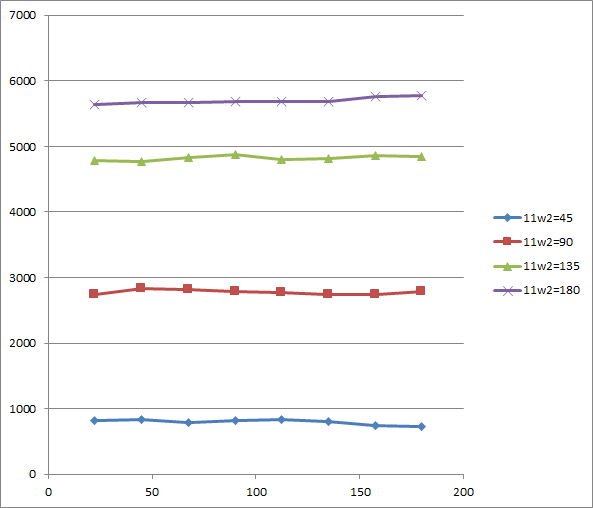} 
\caption{`11' for constant values of $\omega_2$}
\label{fig:subim2}
\end{subfigure}\hfill
\caption{\textbf{Variation of probabilities for different states with $\lambda$ for constant values of $\omega_2$ while $\omega_1$ varies for N=2.} (a) `00'  (b) `01' (c) `10' (d) `11'}
\label{fig15}
\end{figure}

\begin{figure}[H]
\begin{subfigure}{0.5\textwidth}
\includegraphics[width=\linewidth]{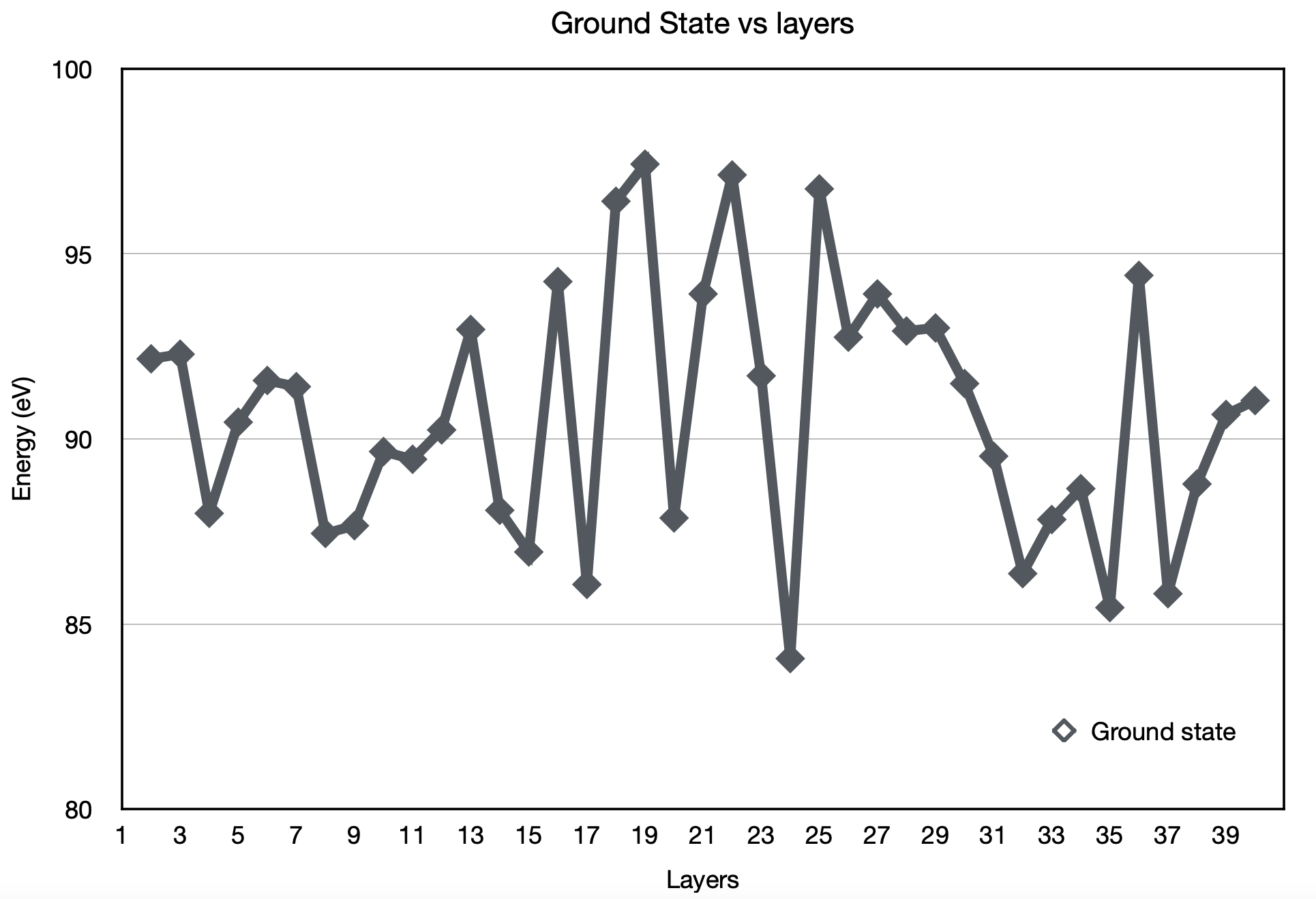} 
\label{fig:subim1}
\caption{\textbf{Variation of Ground State Energy }}
\end{subfigure}\hfill
\begin{subfigure}{0.5\textwidth}
\includegraphics[width=\linewidth]{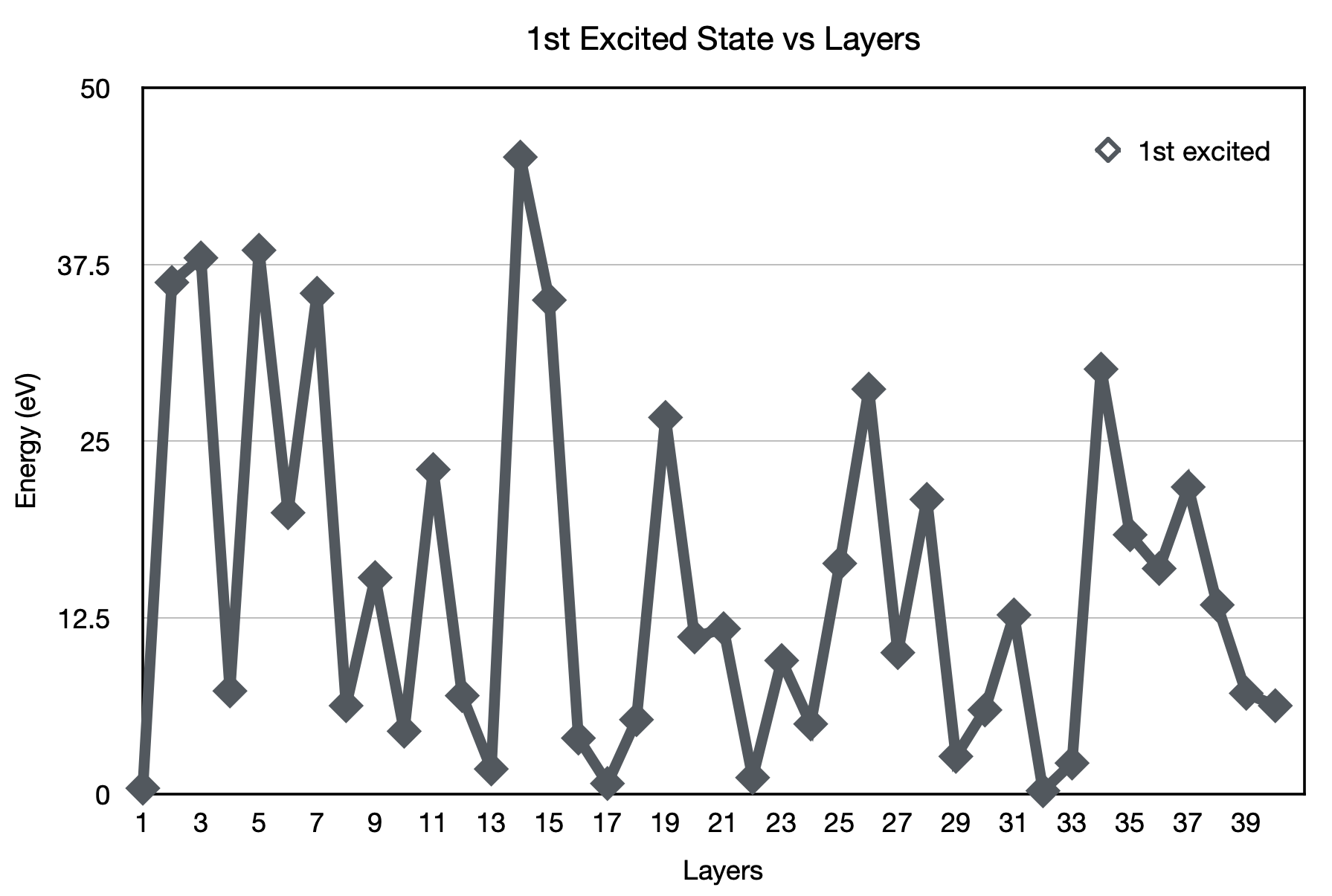} 
\caption{\textbf{Variation of 1st Excited State Energy }}
\label{fig:subim2}
\end{subfigure}\hfill

\caption{\textbf{Variation of both eigensolutions for Ising-Hamiltonian across 40 layers.} (a) Ground State (b) 1st Excited State} 
\label{fig11}
\end{figure}
doesn't restrict the number of achievable states over which the expectation value of $H$ can be taken. In return this ensures that the minimum expectation value is limited only by the capabilities of the classical optimizer.\\

For classical optimizer we have opted to use the Simultaneous Perturbation Stochastic Approximation optimizer (SPSA) optimizer \cite{44}. It is an appropriate optimizer for optimizing a noisy objective function. SPSA approximates the gradient of the objective function with only two measurements. It does so by simultaneously perturbing all of the parameters in a random fashion, rather than the gradient descent where each parameter is perturbed independently. While using VQE in either a noisy simulator \cite{45} or on real hardware, SPSA is used as the recommended classical optimizer.\\

\begin{figure}[H]
\begin{subfigure}{0.5\textwidth}
\includegraphics[width=\linewidth]{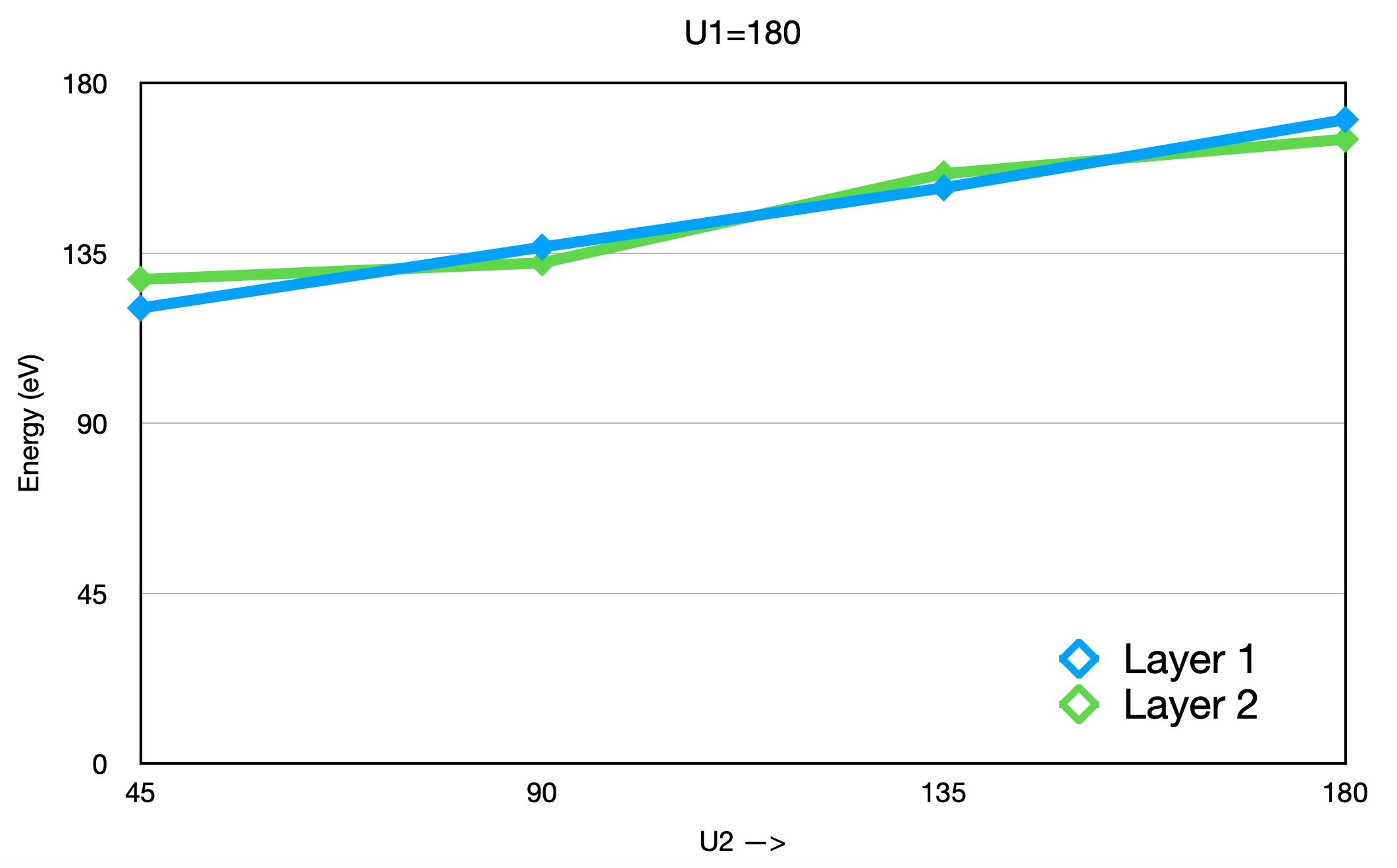} 
\label{fig:subim1}
\caption{\textbf{ $\omega_1$ remains constant  }}
\end{subfigure}\hfill
\begin{subfigure}{0.5\textwidth}
\includegraphics[width=\linewidth]{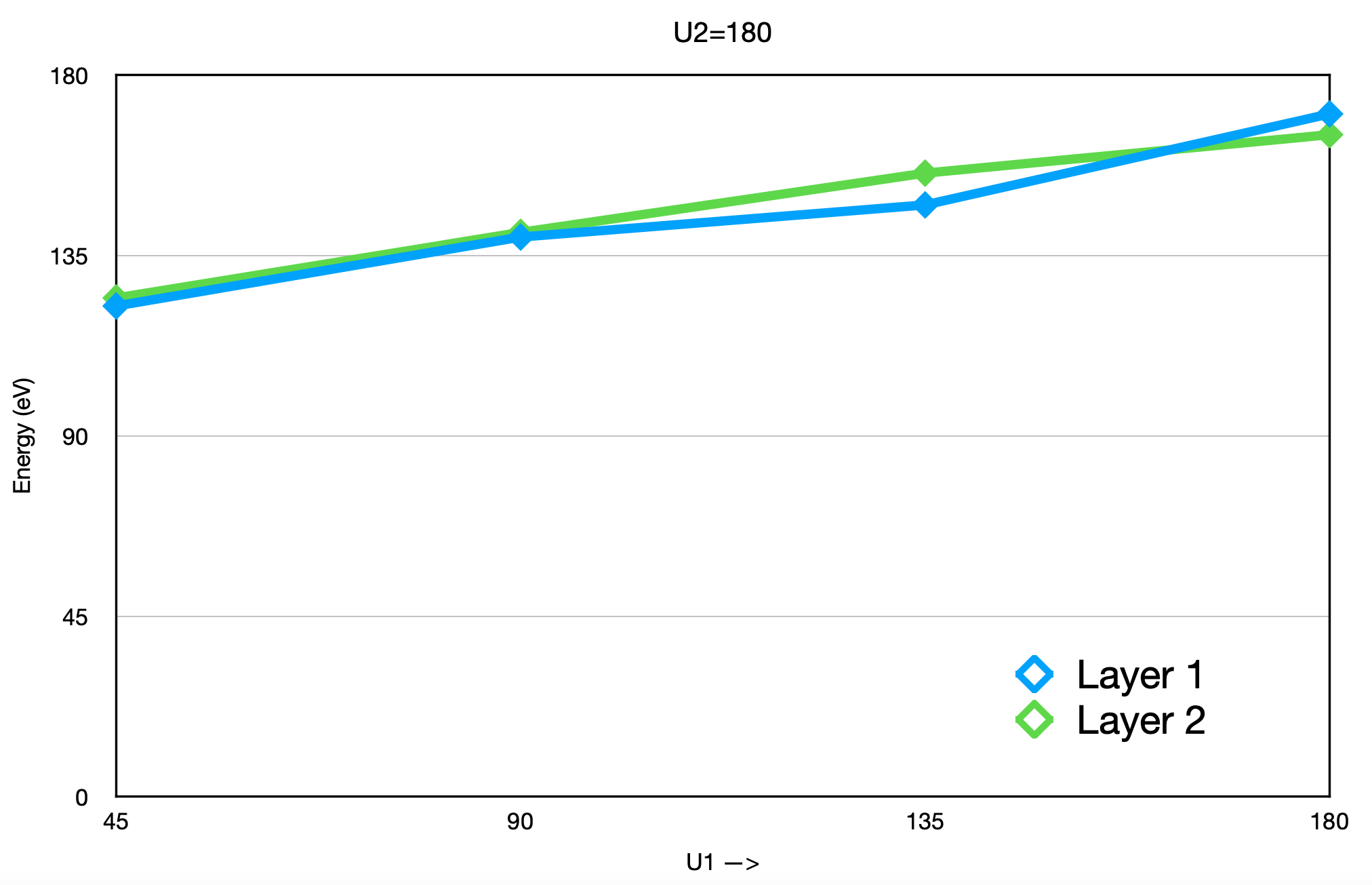} 
\caption{\textbf{ $\omega_2$ remains constant }}
\label{fig:subim2}
\end{subfigure}\hfill
\caption{\textbf{Variation in probabilities of different states with  for constant values of either $\omega$ while  other varies for N=2}. (a) $\omega_1$ constant with U1=$\pi$ (b) $\omega_2$ constant with U2= $\pi$ } 
\label{fig12}
\end{figure}

With the defined ansatz and classical optimizer, we use VQE and obtain both the minimum ground state and 1st excited state energies for various layers of ansatz, all of which are plotted in Fig: \ref{fig11}. Couple of graphs (Fig: \ref{fig12}) comparing ground state energy in different layers with varying $\omega$ have been shown. From Fig: \ref{fig12} no concrete pattern for the peaks and troughs of different energy states with respect to layers could be found.

\section{Fidelity}

\begin{figure}[H]
\centering
\includegraphics[width=0.5\textwidth]{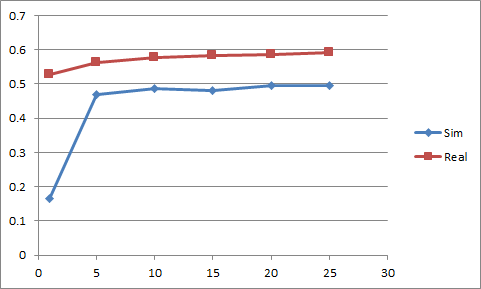}
\caption{Fidelity of real chip (ibmq Santiago) vs simulator of different iterations }
\label{fig13}
\end{figure}

Fidelity $\cite{34}$ is the measure of closeness between two quantum states. In our case, $\psi_\rho$ is the measured state and $\psi_\sigma$ is the pure quantum state. Let's define two density matrices such as $\rho$= $\ket{\psi_\rho}\bra{\psi_\rho}$ and $\sigma$= $\ket{\psi_\sigma}\bra{\psi_\sigma}$ where $\rho $ is the experimental density matrix and $\sigma$ is the theoretical density matrix. So, fidelity is given by $F({\rho, \sigma})$= $\mathrm{Tr}({\sqrt{\sqrt{\rho}{\sigma}{\sqrt{\rho}}}}$). For any $\rho$ and $\sigma$, 0 $\leq$ ${F({\rho, \sigma})}$ $\leq$ 1 and $F({\rho, \rho})$=1. Also the fidelity has symmetric property i.e. ${F({\rho, \sigma})}$= ${F({\sigma, \rho})}$. The measurement is done using X-basis, Y-basis and Z-basis using Santiago chip in IBMQ and QASM simulator for Fig: \ref{fig4}, the measurement of different iterations of above figure is also taken. The results is shown in the given table \ref{tabulu}, and plots in Fig: \ref{fig13}. Fidelity for classical simulator shows better results than the quantum real machine as can be seen from the graph, but the general growth pattern of fidelity is same for both Simulator and quantum real machine.

\begin{table}[]
\centering
\begin{tabular}{|c|c|c|c|c|l|}
\hline
No. of Iteration & Fidelity using  &  Fidelity using \\
 & real chip & Simulator \\
\hline
\hline
1  & 0.1736 & 0.5275  \\ \hline
5  & 0.4723 & 0.5614  \\ \hline
10 & 0.4873 & 0.5759  \\ \hline
15 & 0.4833 & 0.5827  \\ \hline
20 & 0.4942 & 0.5858  \\ \hline
25 & 0.5014 & 0.5915  \\ \hline

\end{tabular}
\caption{Fidelity of real chip vs simulator}
\label{tabulu}
\end{table} 

\section{Conclusion}

We have shown the basic circuit and simulation of the Ising-Hamiltonian which is the sum of all staggered energy and spin coupled complex of a dimer atomic lattice with all the entangled states, for one dimensional dissipative spin chain in magnetic field. This particular case can be used as a test bed to understand steady state dynamics of dissipative spin chains for Hamiltonians of various models such as XXZ model, Heisenberg model \cite{46} etc. We have mentioned about the time correlation factor of the synchronized state which weakly depend upon $\lambda$ which could be calculated by measuring the average value, all the plots signify the time correlation factor and shows the spontaneous synchronization with the resonance peak$ \cite{40}$ and could be further detected the relation between different energy and spin-spin coupling. But there are some points on the graph where each $\omega$ change its phase, where we would like to keep an open question about the behaviour of the $\omega$ in different $\theta$ value of $\lambda t$, which would also depict the relationship between the one-time correlation operator with the synchronization states at particular $\omega$ and $\lambda t$. We have also pointed out the anomalous behaviour of probabilities with respect to constant values of $\omega_1$ which doesn't seem to follow a pattern, which may be used for further studies. We obtained lowest energy values for ground state energy for 24 layer and at 1st layer for first excited state for U3 anstaz while using SPSA optimizer, this gives a possibility of obtaining similar results using different anstaz and optimizers for far less number of layers in case of ground state energy which can be found using further studies of this field.

\section{Acknowledgement}

S.P., A.N., A.M., D.S., T.S.B. and S.K.S would like to thank Bikash's Quantum (OPC) Pvt. Ltd. for providing hospitality during the course of this project. B.K.B. acknowledges IISER-K Institute fellowship. The authors also acknowledge Victor Mukherjee and Rahul Sharma of IISER Berhampur who helped us the peer review respectively. S.P. and A.N. has contributed to the concept, theory, quantum circuit, VQE and the implementation part. D.S. and T.S.B has contributed to the graph plotting. S.K.S has contributed to the entire fidelity section and A.M has helped us in the literature writing of the paper. The authors appreciate IBM quantum experience's assistance in developing the fundamental circuits. The writers' opinions are their own and do not reflect IBM's or the IBM quantum experience team's official position.

\end{document}